# Crossover regimes in lower dimensional structures

R. Batabyal and B. N. Dev*

*Department of Materials Science, Indian Association for the Cultivation of Science,*

*2A and 2B Raja S. C. Mullick Road, Jadavpur, Kolkata 700032, India*

**Abstract**

Modern growth and fabrication techniques can produce lower dimensional structures in the crossover regime. Such structures in the crossover regime can provide tunability of various properties. For example, a zero-dimensional (0-D) structure evolving towards a 2-D structure shows electronic structure which is neither 0-D-like, nor 2-D-like. Within the crossover regime the electronic density of states (DOS) at Fermi level ($E_f$) keeps on changing as the size of the system changes. DOS at $E_f$ determines many properties of materials, such as electronic specific heat, spin susceptibility etc. Keeping the importance of DOS at $E_f$ in mind, we determine their values and other details of electronic structure of lower dimensional structures of metals, in the 0-D to 1-D, 1-D to 2-D, 2-D to 3-D, 0-D to 2-D, 0-D to 3-D and 1-D to 3-D crossover regimes, in a simple free electron model. We compare our results with analytical theory and experimental results, wherever available. We also present some results obtained by scanning tunneling spectroscopy measurements on Ag islands on Si(111) evolving from a 0-D to a 2-D structure. This simple model is quite useful in understanding lower dimensional structures in the crossover regimes.

## I. Introduction

Lower dimensional structures, such as zero-dimensional (0-D), one-dimensional (1-D) and two-dimensional (2-D) structures, have gained importance as the semiconductor fabrication technology has advanced. Practically these structures are quasi-0-D, quasi-1-D and quasi-2-D. Epitaxially grown layers of materials are used for device fabrications. From the originally grown 2-D layers, 1-D and 0-D structures are usually fabricated by lithographic techniquesin the so-called top-down approach. However, for the growth of metallic layers on semiconductor surfaces, which are also technologically important, in many cases a uniform 2-D layer cannot be grown. Very often self-organized compact or elongated islands of deposited materials grow, providing lower dimensional structures.This is the bottom-up approach.Besides the self-organized structures, lower dimensional metal structures, including some structures in the crossover regimes, have also been fabricated by manipulating atoms with a scanning tunneling microscope [1].As far as the thickness or height of an island is concerned, one-atom to a few-atom thick islands with large enough lateral dimensions can be considered 2-D.However, if the lateral dimension of the island is also of this order, the system is to be treated as 0-D instead of 2-D. So, a natural question is: what should be the minimum lateral dimension of such an island, so that it can be considered a 2-D system? Making measurements on 2-D systems are important. Electronic density of states (DOS) of 2-D systems will depend on the thickness, i.e., on the number of atomic layers in the 2-D structure. Electronic DOS at Fermi level of a system determines various properties, such as electronic specific heat, spin susceptibility etc. [2]. As the number of atomic layers in a 2-D system increases, its DOS at Fermi energy would increase and eventually attain the bulk or 3-D value [3]. Thus, all those properties that depend on DOS at Fermi level will vary as the number of atomic layers in the 2-D system varies. Scanning tunneling spectroscopy (STS) measurements on 2-D islands of chosen heights, determined by scanning tunneling microscopy (STM), can reveal electronic structure and properties of such 2-D systems evolving into a 3-D structure. However, to begin with, one needs to know what minimum lateral dimension of an island qualifies for a 2-D system, because if the lateral dimension is too small, the island would be a 0-D system rather than a 2-D system. A simple free electron calculation can provide an idea about this lateral dimension, for which an island can be treated as a 2-D system. Also a guideline can be obtained for other cases: (i) for a 2-D system evolving into a 3-D system, the minimum thickness for qualification as a 3-D system, (ii) for a 0-D system evolving into a 1-D system, the minimum length for the qualification as a 1-D system, (iii) for a 1-D system evolving into a 2-D system, the minimum width for the qualification as a 2-D system and(iv) for a 0-D system evolving into a 3-D system, the minimum volume for the qualification as a 3-D system. In each case, evolution of various properties in the crossover regimes can be estimated.

Various systems in the dimensional crossover regimes have shown interesting results providing a handle on tunability of materials properties. Most often epitaxial growth of Pb on Si leads to island growth [4]. STS measurements of superconducting behaviour on Pb thin films (islands) on Si(111) surfaces forthicknesses ranging from 2 to 5 atoms, i.e., in the 2-D to 3-D crossover regime, have shown the dependence ofthe superconducting transition temperature and the superconducting gap on the island thickness [5]. STS measurements of negative differential resistance (NDR) in electron transport in Ag thin films (islands) on Si(111), for 1 to 5 atomic layers of Ag has shown a thickness dependence of NDR onset voltage [6]. STS experiments on a single Au atom to a linear chain containing up to 15 Au atoms have shown interesting results in the 0-D to 1-D crossover regime [1].

Deposition of about one monolayer (~ $10^{15}$ atoms/cm$^2$)of Ag on Si(111) at room temperature (RT) or at low temperature followed by annealing at RT, shows the growth of Ag islands of 2-atomic layer thickness [7, 8]. If we want to make measurements of electronic DOS on such islands and understand their 2-D properties, what should be the minimum lateral dimension of the island? Definitely a 2-atom thick island with lateral dimensions also containing 2 atoms in each dimension would represent a quantum box or a 0-D system. In order to obtain an idea about the lateral dimensionswhere such Ag islands would behave like a 2-D system,we have carried out a simple free-electron modelcalculation. This simple model also shows dimensional crossover to a 3-D system, as atomic layers are added to a 2-D system.Thicker islands can be grown by depositing additional material, as it was done in Refs. [6, 8].

We begin with a 0-D system where a box contains 8 Ag atoms – 2 atoms each along $x$- ,$y$- , and $z$-direction. Each atom contributing an $s$ electron (Ag $5s$), this quantum box contains 8 electrons. We evaluate its DOS. As expected for a 0-D (or quasi 0-D) system, these are δ-function–like DOS. We then increase only the lateral dimensions ($x, y$) of this structure, thereby increasing the number of atoms or electrons in the structure laterally and follow the evolution of DOS. This way we determine the lateral dimensions at which the DOS attains a clear step-like feature–the signature of a 2-D system. In experiments we expect that an island with these minimum lateral dimensions would display 2-D behavior and hence measurements on such islands can be interpreted as the behavior of a 2-D system. As atomic layers are added on this 2-D island along the z-direction, thereby increasing the island thickness, the evolution of electronic DOS from a 2-D into a 3-D system can be followed.

From the 0-D system, if we increase the dimension in only one direction ($y$), we can follow the evolution of DOS from a0-D to a1-D system. Experimentally this can probably be investigated, as Ag wire-like islands can be grown on a Si substrate [9]. Experiments have been performed on atomic chains of varying lengths within the 0-D to 1-D crossover regime [1]. In order to understand these results a theoretical understanding of evolution of DOS from a0-D to a1-D system is necessary. Next, we fix the dimension of a 1-D island along $y$, and add atomic rows along the $x$-direction and follow the evolution of DOS from the 1-D to a 2-D system. In each case, we determine the Fermi energy ($E_f$) and DOS at $E_f$, as many physical properties of 0-D, 1-D and 2-D systems and those in the crossover regimes would be determined by their DOS at $E_f$.

There is a theoretical model dealing with DOS of quantum dots and crossover from 3-D to 0-D electron gas [10]. This approach does not deal with the situation when atoms, atomic rows or layers are added for the fabrication of 0-D to 3-D (or depleted for 3-D to 0-D) structures. Instead, it keeps the size of the structure fixed and varies the strength of the potential at the boundary of the structure in order to determine the evolution of DOS. This appears to be unrealistic as the potential at the boundary is expected to remain the same for a given material and its surroundings irrespective of its dimension or size. Moreover, it is not easy to determine the DOS at $E_f$ of a system from this model. Our model can be directly applied to experimental situations for lower dimensional systems. DOS for 0-D, 1-D, 2-D and 3-D systems are available in the literature [11, 12]. However, our results additionally provide evolution of several parameters, such as DOS, $E_f$, DOS at $E_f$ and position of the lowest energy state, in all the crossover regimes.

## II. Electronic density of states

Sommerfeld's free electron model describes the motion of electrons confined in a metal. For electrons in a rectangular box of sides $L_x$, $L_y$ and $L_z$ and volume $\Omega = L_xL_yL_z$, the normalized electronic wave function, fulfilling the Dirichlet boundary condition (DBC), $\psi = 0$ on the surface $S$ of the enclosure, has the form [13]

$$\psi(x, y, z) = \left(\frac{8}{L_x L_y L_z}\right)^{1/2} \sin\left(\frac{n_x \pi}{L_x} x\right) \sin\left(\frac{n_y \pi}{L_y} y\right) \sin\left(\frac{n_z \pi}{L_z} z\right) \quad (1)$$

The energy eigenvalues, or the permitted discrete values of $E$, of the electronic states are given by

$$E(n_x, n_y, n_z) = \frac{\hbar^2 \pi^2}{2m}\left(\frac{n_x^2}{L_x^2} + \frac{n_y^2}{L_y^2} + \frac{n_z^2}{L_z^2}\right) \quad (2)$$

where $\hbar$ is Planck's constant, $m$ is the mass of an electron and

$$n_{x,y,z} = 1, 2, 3, 4, \ldots \quad (3)$$

are quantum numbers [13].

The electronic density of states (DOS), i.e., the number of states lying between the energies $E$ and $E+dE$, per unit volume for a bulk ($\Omega \to \infty$) three-dimensional (3-D) material is given by

$$\rho^{3-D} = \frac{(2m/\hbar^2)^{3/2}}{2\pi^2} E^{1/2} \quad (4)$$

Electron mass, $m$, is usually replaced by an effective mass $m_c$. Then the DOS for 3-D and other lower dimensional systems, such as 2-D, 1-D and 0-D systems is given by the well-known analytical expressions [11]

$$\rho^{3-D}(E) = \frac{(2m_c/\hbar^2)^{3/2}}{2\pi^2} E^{1/2} \quad (5)$$

$$\rho^{2-D}(E) = \frac{m_c}{(\pi \hbar^2 L_z)} \sum_{n_z} H[E - E_{n_z}] \quad (6)$$

$$\rho^{1-D}(E) = \frac{(m_c/2\hbar^2)^{1/2}}{(\pi L_y L_z)} \sum_{n_y, n_z} [E - E_{n_y} - E_{n_z}]^{-1/2} \quad (7)$$

$$\rho^{0-D}(E) = \frac{1}{(L_x L_y L_z)} \sum_{n_x, n_y, n_z} \delta[E - E_{n_x} - E_{n_y} - E_{n_z}] \quad (8)$$

where $\delta$ is the Dirac $\delta$-function and $H$ is the Heaviside step function. The DOS ($\rho$) is evaluated at energy $E$. We will work with the 0-D case, and as we will see, all other cases will evolve from the 0-D case. Taking the two-fold spin degeneracy into account and using the Gaussian representation of the $\delta$-function, we can write Eq. (8) as

$$\rho^{0-D}(E) = \underset{\sigma \to 0}{Lt} \sqrt{\frac{2}{\pi \sigma^2}} \left(\frac{1}{L_x L_y L_z}\right) \sum_{n_x, n_y, n_z} \exp\left(\frac{-\left[E - \{A_x n_x^2 + A_y n_y^2 + A_z n_z^2\}\right]^2}{2\sigma^2}\right) \quad (9)$$

where $A_x = \frac{\pi^2 \hbar^2}{2m_x L_x^2}$, $A_y = \frac{\pi^2 \hbar^2}{2m_y L_y^2}$ and $A_z = \frac{\pi^2 \hbar^2}{2m_z L_z^2}$.

with $m_x$, $m_y$ and $m_z$ being the effective mass of electron for motions along x-, y- and z-direction, respectively. We will use $m_x = m_y = m_z = m = 9.109 \times 10^{-31}$ kg.

For the periodic boundary condition (PBC)

$n_{x,y,z}$ = ……, -4, -2, 0, +2, +4, …. (10)

For small systems PBC is not appropriate [3]. We use DBC for all our computations. We use $n_{x,y,z}$=1, 2, 3 … 1000. The value of σ is taken as 0.1 eV.

We will begin with the 0-D case (quantum box) and simulate the evolution of DOS from 0-D to all other dimensions. This can be done from Eq. (8) or Eq. (9).It is not necessary to use the other analytical forms [Eqs. (5) – (7)] of DOS.

**A. 0-D to 2-D**

Now let us take a metal structure, where in all the directions electrons are confined. We have considered a quantum box with height $L_z$, length $L_y$ and width $L_x$ ($L_x=L_y= L$). The density of states of such a confined system will be

$$\rho(E) = \underset{\sigma \to 0}{Lt} \sqrt{\frac{2}{\pi\sigma^2}} \left(\frac{1}{L^2 L_z}\right) \sum_{n_x,n_y,n_z} \exp\left(\frac{-\left[E - \left\{A_z n_z^2 + A_p\left(n_x^2 + n_y^2\right)\right\}\right]^2}{2\sigma^2}\right) \quad (11)$$

where $A_p = \frac{\pi^2 \hbar^2}{2mL^2}$

We begin with a quantum box that contains 8 atoms, 2 atoms in each direction. The typical size of this box is $L_x=L_y = L_z = 0.5$ nm. We make this choice based on the experimental observation [7, 8] of Ag growth on Si(111). Ag initially grows as a 2-atom thick island with (111) crystallographic orientation [Ag(111)/Si(111)]. In fcc bulk Ag, (111) planner spacing is 0.23 nm. So the height of a 2-atom thick island would be 0.46 nm. However, in experiments this value has been found to be 0.54 nm [7, 8]. As the island thickness increases, this planer spacing tends to reach its bulk value [14]. For calculation we have taken the height of this 2-atom thick island to be 0.5 nm. So, an island of $(0.5)^3$ nm$^3$ volume contains 8 Ag atoms and each atom contributes an *s* electron (Ag *5s*). (We ignore any contribution from the *d* electrons).With each atom contributing one electron, this box would contain 8 electrons. The DOS calculated from Eq. (11) for this box is shown in the lowest panel of Fig. 1. As expected, DOS contains δ-function-like discrete peaks. Now keeping the value of $L_z = 0.5$ nmfixed, we increase the lateral dimensions ($L_x=L_y=L$). The evolution of DOS with the increase of the value of *L* is seen in Fig. 1. For *L*=5nm, the 2-D feature of step function behavior in DOS is already seen with some wiggles on the steps. For *L*=15 nm smooth step features of expected DOS of a 2-D system [Eq. (6)] are observed. So, it may be concluded that a 2-atom thick island with a lateral dimension of ~ 15 nm or about 60 atoms would suffice to be considered a 2-D system. Results of measurements of some parameters on such an island could be interpreted as their value for the 2-D form of that material.

**B. 2-D to 3-D**

From Fig. 1 it is seen that a lateral dimension of 15 nm ($L_x=L_y$=15 nm) represents a 2-D system. Now we add atomic layers on this 2-D system to approach a 3-D system. A thickness of 1 nm would contain 4 atomic layers of Ag. Fig. 2 shows the evolution of DOS as the layer thickness is increased. We notice that a 2 nm thick 2-D island, i.e., an eight-atom thick island, already shows the trend of the 3-D characteristics of $E^{1/2}$ dependence of DOS, albeit with some wiggles on DOS. When the thickness is increased to 10 nm a smooth $E^{1/2}$ dependence of DOS, typical for a 3-D system [Eq. (1)], is observed. Investigation of superconducting behavior, such as superconducting transition temperature and superconducting gap, on Pb islands on Si(111), has revealedtheir evolution in 2-5 atomic layer thick islands [5]. Also in Ag islands on Si(111), negative differential resistance (NDR) behavior in electron tunneling conductance has shown an evolution as the thickness of the island has varied from 1 to 5 atomic layers of Ag [6]. In Ref [6] ab-initio density functional theory (DFT) calculations, showing the evolution of DOS with thickness, have been used to interpret the evolution of NDR.DFT calculations have shown that the energy positions of specific peaks in DOS shift to lower energies with the increasing number of atomic layers. This feature has been observed in experiments on 4 and 5 atomic layer thick Pb films on Si(111) [5]. As we will see latter (Section III) that our simple free electron calculation also shows this shift in energy position with increasing layer thickness.

## C. 0-D to 1-D

From the 0-D system ($L_x=L_y=L_z=0.5$ nm) we gradually increase the value of $L_y$ keeping dimensions in the $z$- and $x$-direction fixed ($L_x=L_z=0.5$ nm). The DOS is shown in Fig. 3 for increasing values of $L_y$. We notice that for $L_y=7$ nm, the features of 1-D DOS can be easily identified, albeit with some wiggles, which are remnants of 0-D features. For $L_y=15$ nm, clear 1-D features, as expected from Eq. (7), are observed.

We provide a qualitative comparison of our results with the experimental results of Nilius *et. al.*[1]. These are scanning tunneling spectroscopy (STS) experiments on Au atomic chains on NiAl substrates. Conductivity, which is proportional to DOS, vs sample bias are presented in Fig. 2 of Ref. 1. As the length of the chain increases from one atom to a few atoms, the conductivity peak at the lowest energy shifts to lower energy. The magnitude of this shift gradually decreases as the chain length increases. From 15 atoms ($Au_{15}$) onwards, this shift is negligible. This trend is seen from Fig. 3. From Fig. 3 and also from additional calculations for intermediate lengths, we plot the energy of the lowest energy peak vs length of the structure in the 0-D to 1-D crossover regime in Fig. 4. We notice fromFig. 4 that from a length of 4 nm (or 16 atoms) onwards, practically there is no shift in energy of the lowest energy peak position. The DOS for $Au_{20}$/NiAl, presented in Fig. 4(D) in Ref. 1 shows wiggles on the $E^{-1/2}$ decay of the 1-D structure [Eq. (7)]. This means that even for this length, DOS still displays the 0-D remnant and the 1-D feature is not fully developed. DOS for such structures (for $L_y$ = 4 nm) and for fully developed 1-D structures ($L_y$ = 15 nm) are shown in Fig. 5, which also shows that the shift in energy position of the lowest energy peak is insignificant for these cases.

## D. 1-D to 2-D

Beginning from the DOS at the topmost panel ($L_x = L_z = 0.5$ nm, $L_y = 15$ nm) in Fig. 3, we add atomic rows along the $x$-direction, i.e., we increase the value of $L_x$. We show the evolution of DOS for this case in Fig. 6. For $L_x=7$nm, DOS already reveals 2-D features with some remnant wiggles of the 1-D features. The topmost panel (for $L_x=L_y=15$ nm, $L_z=0.5$ nm) shows the DOS, typical of a 2-D system [Eq. (6)]. This system contains 2 atoms in the z-direction and about 60 atoms each along $x$- and $y$-direction.

## E. 0-D to 3-D

For quantum dots growing in size (volume), evolution from 0-D to 3-D is relevant. Here the size of the system ($L$) is increased keeping $L_x = L_y = L_z = L$. The evolution of DOS with the system size is shown in Fig. 7. For $L = 3$ nm the system has already developed the 3-D feature, albeit with some remnants of the 0-D features. For $L = 10$ nm, fully developed 3-D features, i.e. $E^{1/2}$ dependence of DOSis observed. This case is applicable for metal nanoparticles with increasing diameter.

## F. 1-D to 3-D

This is a case relevant for a nanowire growing in lateral size or diameter.The results for this case are shown in Fig. 8.Calculations are for$L_y$ = 15 nm and $L_x = L_z= L$. For $L = 2$ nm, the 3-D behavior has already developed along with some remnants of the 1-D features of DOS. *Ab initio* calculations based on density-functional theory for freestanding Na (free-electron-like) nanowires, with increasing cross section of the nanowire, shows similar features [15] as in Fig.8. In the calculations in Ref. [15], a 10-atom wide wire shows the development of 3-D features as in the case of $L_x = L_z = 2$ nm (or 3 nm) in Fig. 8, or equivalent to a 8-atom (12-atom) wide wire.

## III. Lowest energy position

In the dimensional crossover regimes, specific features in DOS shift to lower energies as the system size increases. Here we discuss the shift of only the lowest energy feature in the DOS. As mentioned earlier, in Fig. 4 we have shown the energy variation of the lowest energy feature in DOS for a structure evolving from 0-D to 1-D. In Fig. 4 we also show this variation for structures evolving from 0-D to 2-D, 1-D to 2-D, 2-D to 3-D, 0-D to 3-D and 1-D to 3-D for comparison. As seen from the figure, in all cases, this variation is strong only at the beginning of the evolution in the crossover regime. (Note that for the 0-D to 2-D evolution, the abscissa in Fig. 4, represents the value of both $L_x$ and $L_y$, $L_x = L_y = L$, for 1-D to 3-D evolution the abscissa represents $L_x = L_z = L$ and for 0-D to 3-D evolution, the abscissa represents all three dimensions $L_x = L_y = L_z= L$). For 2-D to 3-D evolution, this feature has been observed experimentally [5] for Pb films on Si(111),where comparison of energy positions for specific peaks for 4 and 5 atomic layer thick films shows this trend. Also for Ag on Si(111),  this feature has been revealed and interesting thickness dependent NDR has been observed [6]. Using a scanning tunneling microscope, appropriate samples can be prepared as well as investigated. Layer thickness of a film (or flat top island) can be increased by one atomic layer by applying a voltage pulse from a tunneling microscope [16]. Successive pulses would increase the thickness by one atomic layer for each pulse. So, one would obtain samples over a wide range of thickness. The energy positions of specific peaks can be easily investigated with the same microscope by scanning tunneling spectroscopy experiments.

### IV. Fermi energy and DOS at Fermi energy

As the total number of electrons in each system is known, Fermi energy is obtained in each case by arranging the eigenvalues in ascending numerical order, and then filling successive energy levels from the bottom and obtaining the highest filled energy level. Fermi energies ($E_f$) for 0-D to 1-D, 1-D to 2-D, 0-D to 2-D, 2-D to 3-D, 0-D to 3-D and 1-D to 3-D cases are shown in Fig. 9. Once the value of $E_f$ is known, Eq. (9) is used to obtain DOS at $E_f$. Fig. 10 shows the corresponding DOS at $E_f$.

Ref.[3] has dealt with quantum size effects in the Fermi energy and electronic density of states in metallic thin films. The dependence of $E_f$ and DOS at $E_f$ on film thickness has been worked out in a free electron model. We show a comparison of our results of DOS at $E_f$ for the 2-D to 3-D crossover regime [Fig. 10(d)]. In order to compare with the results in Ref. [3] we carried out the calculation in small steps of thickness. The results are in good agreement with those in Ref. [3]. In Ref. [3] periodic boundary conditions (PBC) have been used in $x$- and $y$-direction, as these dimensions are large, while Dirichlet boundary condition (DBC) has been used in the $z$- direction. In Fig. 10(d) we have shown results using DBC in all directions. However, we also carried out computation with the boundary conditions as in Ref. [3]. The results are practically the same. When we carried out computations for $E_f$ at small thickness steps as in Ref. [3], we found oscillatory decrease of the value of $E_f$ with thickness, as in Ref. [3]. (These results are not shown here).The oscillatory behavior is not seen for $E_f$ in Fig. 9(d), where the thickness steps are large.

### V. DOS at $E_f$ for evolution from 2-D to 3-D: Comparison with analytical theory

The periodic boundary condition, used for a bulk material, is not appropriate for an ultrathin film. For an ultrathin film, Dirichlet boundary condition is appropriate and the consequence of that is reflected in the value of DOS at $E_f$. DOS at $E_f$ for an ultrathin film (2-D) is given by [3]

$$\rho(E_F) = \rho^0(E_F^0)\left[1 - \frac{\gamma^0}{L_z}\right] \quad (12)$$

where, $\gamma^0 = \frac{\pi}{4k_F^0} - \frac{1}{2k_F^0}\left\{(2+\lambda^0)\sin^{-1}\left(\frac{1}{\sqrt{\lambda^0}}\right) - (\lambda^0 - 1)^{\frac{1}{2}}\right\}$ and $\rho^0(E_F^0)$ is DOS at $E_f$ for a bulk (3-D) system.

Here, $k_F^0$ is the bulk Fermi wave vector, $\lambda^0 = \left(\frac{k_{TOP}}{k_F^0}\right)^2$ with $k_{TOP} = \frac{\sqrt{2mV_z}}{\hbar}$, $V_z$ is the square well potential depth and $m$ is electron mass.

DOS at $E_f$, normalized to its bulk (or 3-D) value is given by

$$\frac{\rho(E_F)}{\rho^0(E_F^0)} = 1 - \frac{\gamma^0}{L_z} \quad (13)$$

Eq. (13) shows the dependence of DOS at $E_f$ on $L_z$, and hence 2-D to 3-D evolution of DOS at $E_f$. The average DOS at $E_f$ from our free electron model (Fig. 10(d)) shows dependence on film thickness similar to that in Eq. (13).

## VI. Experimental Results

We present some limited experimental results here. As the Ag/Si(111) system shows the growth of two-atom thick islands of various lateral dimensions, we try to obtain some data for the 0-D to 2-D evolution. The experiments were performed using a compact molecular beam epitaxy (MBE) system with an attached variable temperature scanning tunneling microscope (VT-STM) [Omicron Nanotechnology], operating under ultrahigh vacuum (UHV) condition. The base pressure in both the MBE growth chamber and the STM chamber was less than $1.0 \times 10^{-10}$ mbar. N-type Si(111) substrates were degassed at $600^\circ C$ for 15 hrs. and flashed at around $1200^\circ C$ for a minute to obtain clean 7×7 reconstructed surfaces. The growth of Ag on Si(111)-7×7 surfaces is achieved by depositing Ag atoms onto the substrates at room temperature (RT), from a Knudsen cell maintained at 830ºC. Initially we observed the reconstructed Si(111)-(7×7) surface and the morphology of the Ag films using a clean W tip in scanning tunneling microscopy (STM). Electronic DOS at Fermi level was determined by *current (I) – voltage (V)* measurement in scanning tunneling spectroscopy (STS). A sub-monolayer (ML) Ag film was deposited on a Si(111)-(7×7) substrate. [Here 1 ML of Ag is equivalent to the nominal surface atomic density of Ag(111), $1.5 \times 10^{15}$ atoms/cm$^2$, as in Ref. [7, 8]]. Low coverage of Ag on Si (111)-7×7 surfaces offers only two atomic layer thick (0.54 nm) islands with various lateral dimensions (areas). This strong height selectivity allows us to investigate the DOS of the islands with a fixed height but different areas. Thus, 0-D to 2-D evolution of DOS can be determined from experiments. Numerical derivatives, *dI/dV*, were obtained from the *I-V* data. Then the value of *(dI/dV)/(I/V)* at *V = 0* provides the DOS at $E_f$. Fig. 11 shows a STM image, revealing Ag islands, and the DOS at $E_f$ for some small islands. The trend of evolution is in reasonable agreement with the free electron model, Fig. 11(c).

The details in Fig. 10(c) [or 11(c)] for small sizes cannot be directly compared with the experimental data for several reasons. The fluctuation in DOS depends on the actual size of the island. For the 8 atom 0-D system, if $L_z$ = 0.54 nm and $L_x = L_y$ = 0.576 nm are taken instead of $L_x = L_y = L_z$ = 0.5 nm, the fluctuation on data points for small sizes changes quite a bit. The choice of $L_x = L_y$ = 0.576 nm is because of coincidence site lattice matching epitaxy of Ag(111) on Si(111), where 4 unit cells of Ag match with 3 surface unit cells of Si(111) [17, 18]. The change of the values of $L_x$, $L_y$ and $L_z$ amounts to change of electron density, which affects DOS. Moreover, fluctuation of DOS at $E_f$ is stronger at small sizes, as evident from Fig. 10(d) and other panels in Fig. 10. The detailed comparison is not possible between experiment and the free electron model also for other reasons, such as, deviation of effective mass from the free electron mass and the proximity effect. The electronic DOS in the Ag system is influenced by the proximity of Si across the interface [18]. Also Ag induces metal induced gap (MIG) states within the band gap of Si [18]. However, our free electron model appears to agree with the observed trend in the 0-D to 2-D crossover regime. This trend is shown by the smooth line in Fig. 11(c).

## VII. Conclusions

We have investigated the electronic structures in the crossover regimes in lower dimensional structures within a free electron model. For the 0-D to 1-D, 1-D to 2-D, 2-D to 3-D, 0-D to 2-D, 0-D to 3-D and 1-D to 3-D crossover regimes, electronic density of states (DOS), lowest energy peak position, Fermi energy ($E_f$) and DOS at $E_f$ have been determined. For ultrathin films, analytical theory for the dependence of DOS at $E_f$ on the film thickness is available in the literature. Our evaluated DOS at $E_f$ for the 2-D to 3-D crossover regime agrees well with the analytical theory. The trend of variation of the energy position of the lowest energy peak (or any specific peak) for the 0-D to 1-D and 2-D to 3-D regimes, where experimental results are available, compares well with experiments as well as with density functional theory (DFT) calculations. We have also presented some experimental results in the 0-D to 2-D crossover regime, obtained by scanning tunneling spectroscopy, for Ag islands epitaxially grown on Si(111). The trend of variation of DOS at $E_f$ agrees well with the free electron model calculation. We hope that these results would be helpful in providing a guideline for an understanding of lower dimensional systems in the crossover regimes. As various structures in the dimensional crossover regimes can indeed be fabricated, tunability of materials properties can be achieved in such structures.

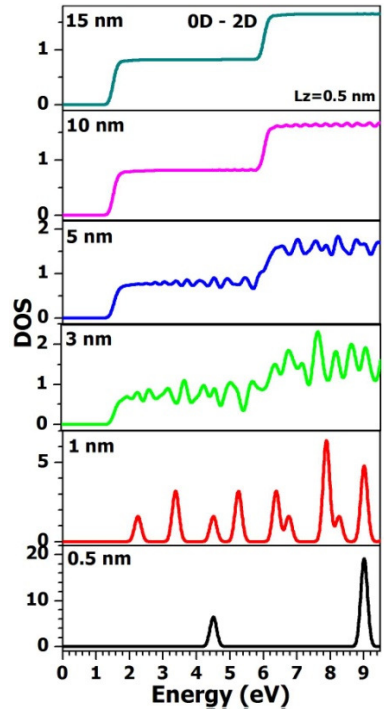

**Figure 1:** Calculated DOS (Number of states × $10^{-1}$/eV-nm$^3$) and its evolution from a 0-D to a 2-D system for increasing lateral dimensions. Upper left corner of the each panel shows the lateral dimensions ($L_x = L_y = L$) and the thickness $L_z$ is fixed at 0.5 nm.

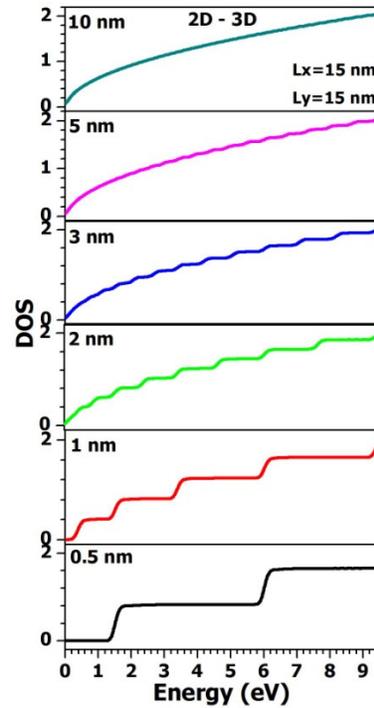

**Figure 2:** Calculated DOS (Number of states × $10^{-1}$/eV-nm$^3$) and its evolution from a 2-D to a 3-D system for increasing thickness. Upper left corner of the each panel shows the thickness ($L_z$) and the lateral dimensions ($L_x = L_y = L$) are fixed at 15 nm for each case.

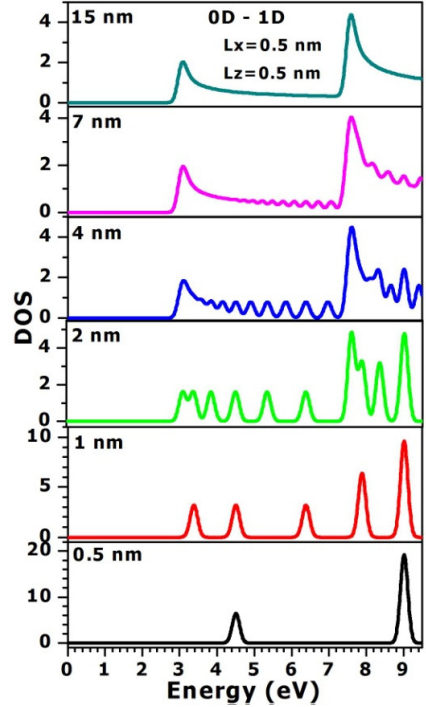

**Figure 3:** Calculated DOS (Number of states × $10^{-1}$/eV-nm$^3$) and its evolution from a 0-D to a 1-D system for increasing length. Upper left corner of the each panel shows the length ($L_y$) and the other two confined dimensions ($L_x = L_z = L$) are fixed at 0.5 nm for each case.

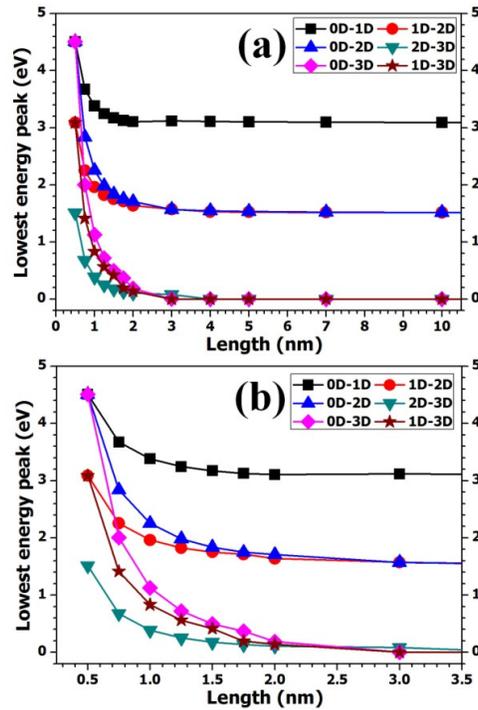

**Figure 4: (a)** The energy variation of the lowest energy feature in DOS for the structures evolving from 0-D to 1-D, 1-D to 2-D, 0-D to 2-D, 2-D to 3-D, 0-D to 3-D and 1-D to 3-D. For the 0-D to 2-D evolution, the abscissa represents the value of both $L_x$ and $L_y$, $L_x = L_y = L$, for 1-D to 3-D evolution the abscissa represents $L_x = L_z = L$ and for 0-D to 3-D evolution, the abscissa represents all three dimensions $L_x = L_y = L_z = L$. **(b)** Expanded view of (a) for the small length region.

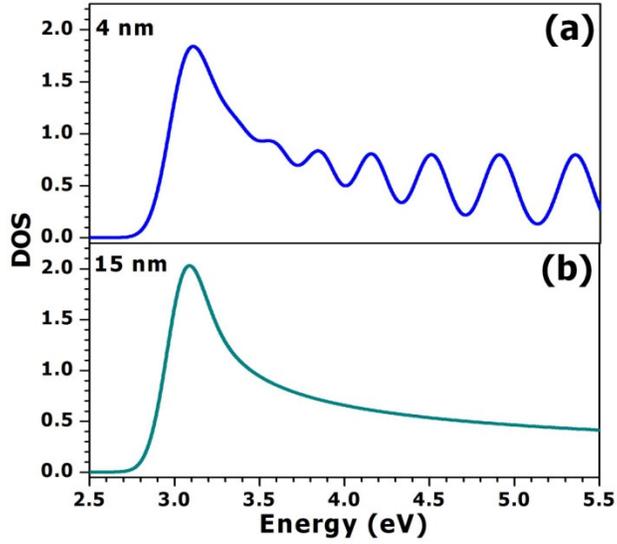

**Figure 5:** A comparison between DOS (a) for a structure in the 0-D to 1-D crossover regime approaching 1-D (for $L_y$ = 4 nm), where 0-D features are still observed, and (b) for a fully developed 1-D structure (for $L_y$ = 15 nm).

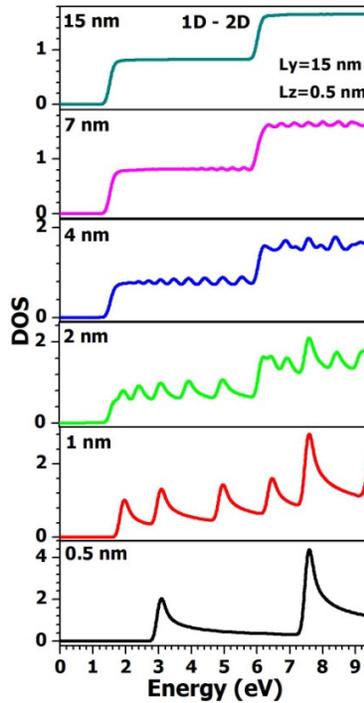

**Figure 6:** Calculated DOS (Number of states × $10^{-1}$/eV-nm$^3$) and its evolution from a 1-D to a 2-D system for increasing width. Upper left corner of the each panel shows the value of $L_x$. The other two dimensions $L_y$ (length) and $L_z$ (thickness) are fixed at 15 nm and 0.5 nm, respectively.

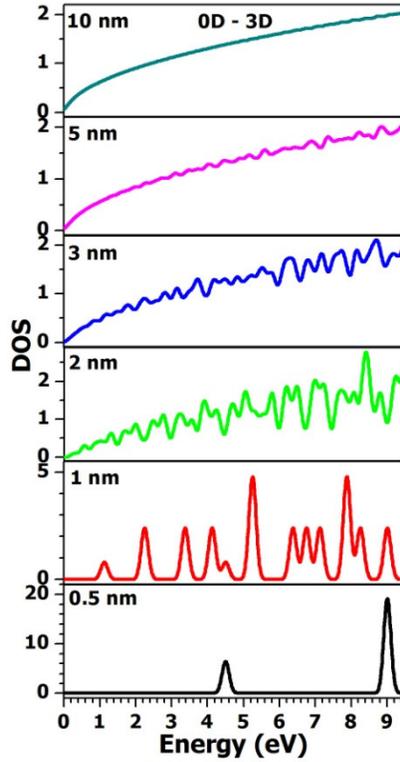

**Figure 7:** Calculated DOS (Number of states × $10^{-1}$/eV-nm$^3$) and its evolution from a 0-D to a 3-D system for increasing volume. Upper left corner of the each panel shows the dimension ($L$) of the cubic box ($L_x = L_y = L_z = L$).

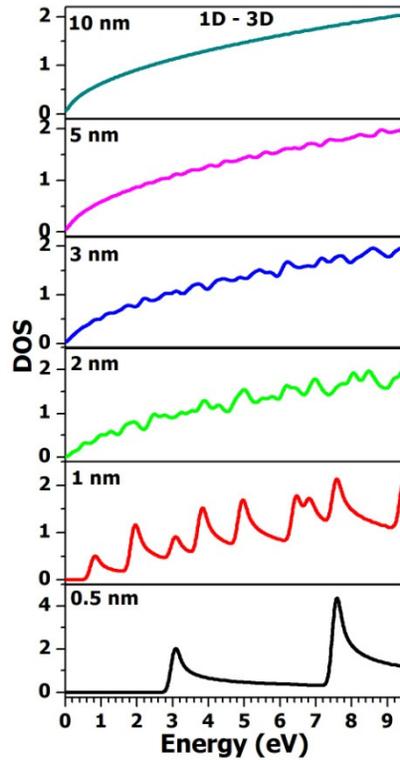

**Figure 8:** Calculated DOS (Number of states × $10^{-1}$/eV-nm$^3$) and its evolution from 1-D to 3-D. Upper left corner of each panel shows the dimensions ($L_x = L_z = L$) and $L_y$ (length) is fixed at 15 nm.

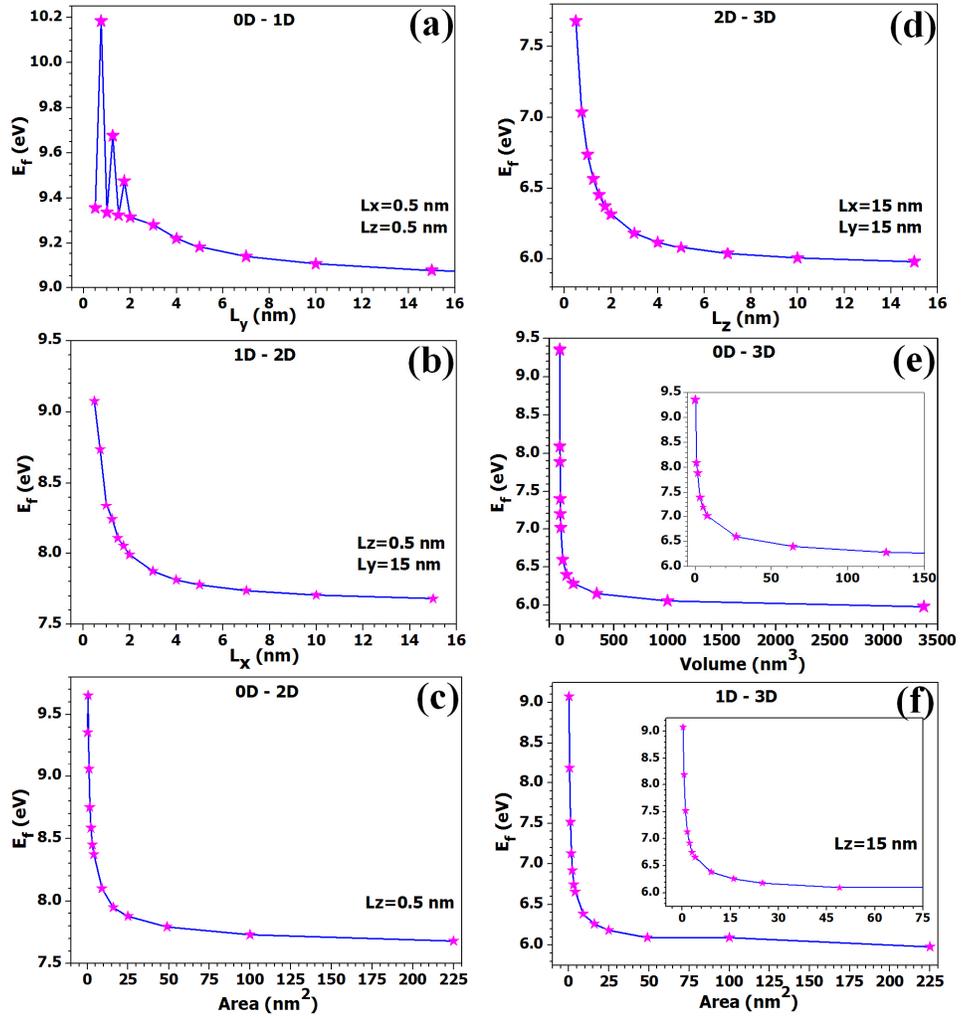

**Figure 9:** Fermi energies ($E_f$) are plotted for the structures evolving from (a) 0-D to 1-D, (b) 1-D to 2-D, (c) 0-D to 2-D, (d) 2-D to 3-D, (e) 0-D to 3-D and (f) 1-D to 3-D systems. The insets show the crossover regime more clearly.

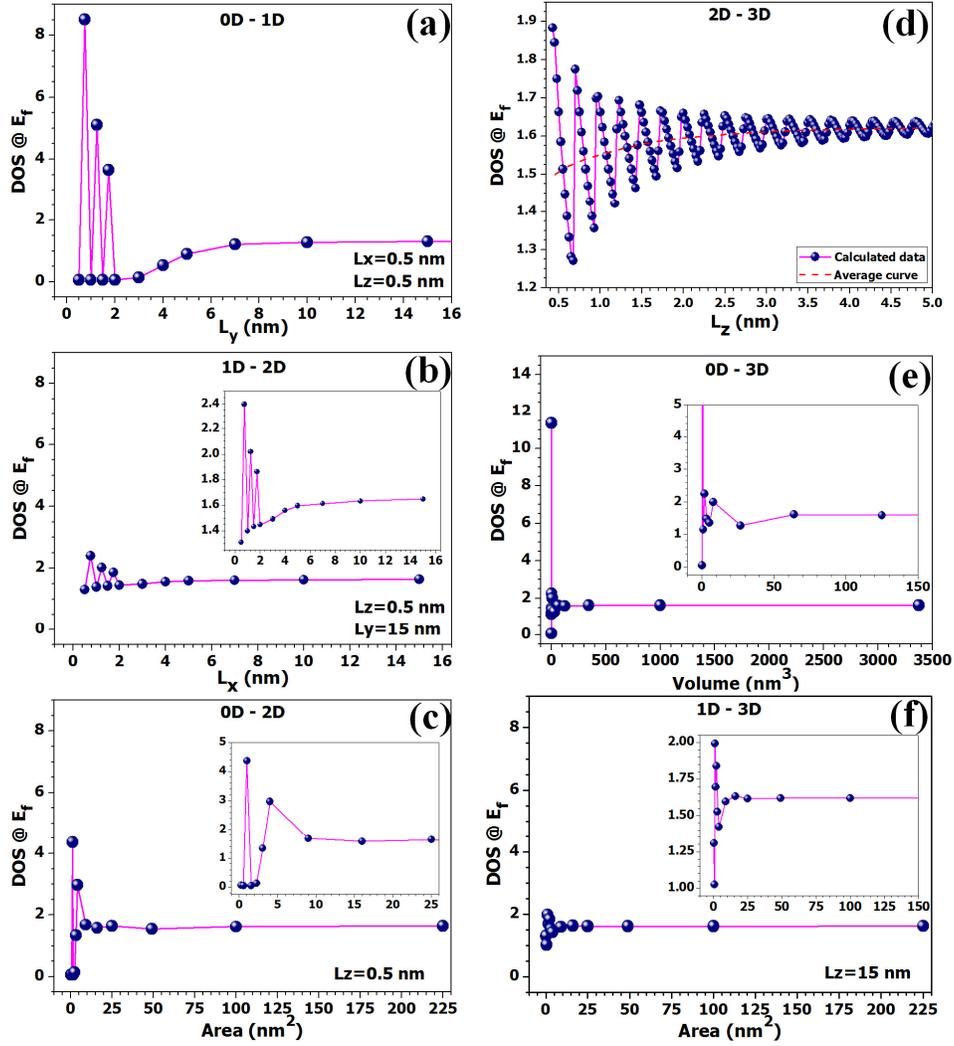

**Figure 10:** DOS at Fermi energy for the structures evolving from (a) 0-D to 1-D, (b)1-D to 2-D, (c) 0-D to 2-D, (d) 2-D to 3-D, (e) 0-D to 3-D and (f) 1-D to 3-D systems. The insets show the crossover regime more clearly. Fluctuations in the value of DOS at $E_f$ are, in general, larger for small sizes.

.

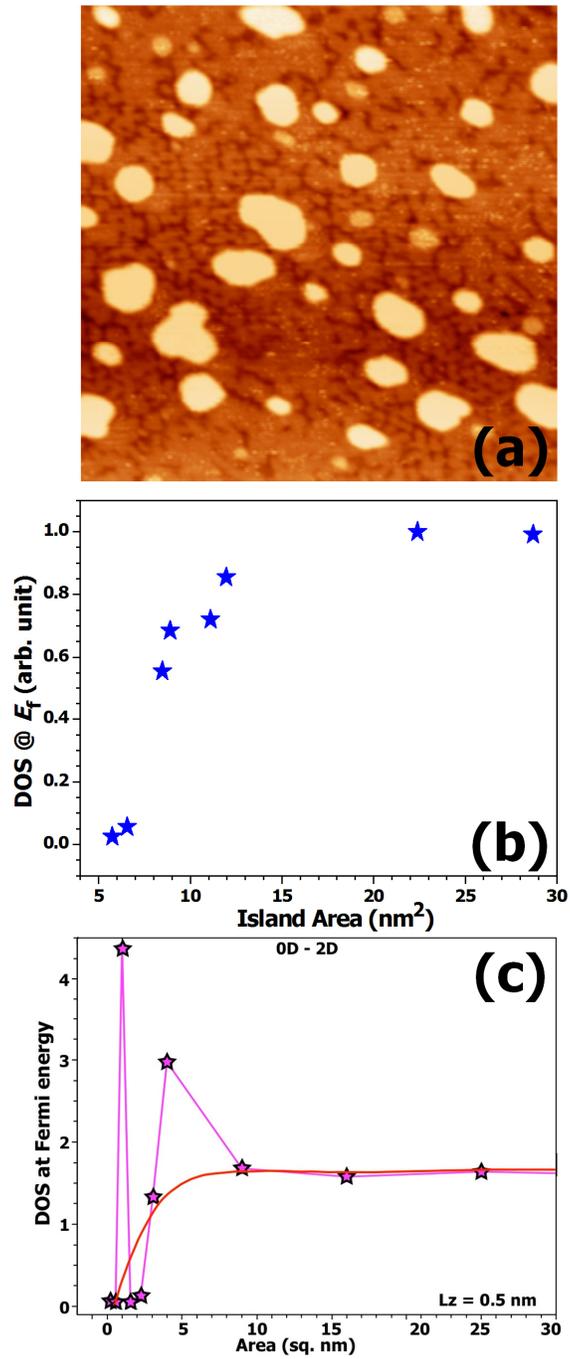

**Figure 11:** (a)A constant current STM image (100 nm × 100 nm) from 0.7 ML Ag, deposited on a Si(111)-7×7surface. All Ag islands are two-atomic layer thick (0.54 nm) but with varying areas. (b) DOS at $E_f$ from islands of selected small areas. (c) calculated DOS at $E_f$ (as in 10(c)) with an average curve (red line) showing the trend..

**References:**

[*]Email: msbnd@iacs.res.in

**Acknowledgement:**

We thank Prof. S. D. Mahanti for suggesting the free electron calculation, while discussing some of our experimental results, when one of us (BND) visited Michigan State University, East Lansing. We also thank Prof. K. Sengupta of our institute for helpful discussions.